# Rapid infrared spectroscopic nano-imaging with nano-FTIR holography

Martin Schnell[1]*, Monika Goikoetxea[1], Iban Amenabar[1], P. Scott Carney[2], and Rainer Hillenbrand[3,4]

1 CIC nanoGUNE BRTA, 20018 Donostia-San Sebastián, Spain

2 The Institute of Optics, University of Rochester, 480 Intercampus Drive, Rochester, NY 14620

3 IKERBASQUE, Basque Foundation for Science, 48013 Bilbao, Spain

4 CIC nanoGUNE BRTA and Department of Electricity and Electronics, UPV/EHU, 20018 Donostia-San Sebastián, Spain

* schnelloptics@gmail.com

**Abstract**

**Scattering-type scanning near-field optical microscopy (s-SNOM), and its derivate, Fourier transform infrared nanospectroscopy (nano-FTIR) are emerging techniques for infrared (IR) nanoimaging and spectroscopy with applications in diverse fields ranging from nanophotonics, chemical imaging and materials science. However, spectroscopic nanoimaging is still challenged by the limited acquisition rate of current nano-FTIR technology. Here we combine s-SNOM, nano-FTIR and synthetic optical holographic (SOH) to achieve infrared spectroscopic nanoimaging at unprecedented speed (8 spectroscopically resolved images in 20 min), which we demonstrate with a polymer composite sample. Beyond being fast, our method promises to enable nanoimaging in the long IR spectral range, which is covered by IR supercontinuum and synchrotron sources, but not by current quantum cascade laser technology.**

**Introduction**

Scattering-type scanning near-field optical microscopy (s-SNOM) [1] is an emerging tool for infrared (IR) nanoimaging that extends the analytical power of infrared light to the deep subwavelength scale. It has thus found successful application in a variety of fields, including the nanoscale identification of molecular vibration signatures [2–6], the probing of polaritonic excitations in novel nanophotonic structures [7–10], the mapping of the free carrier concentration in semiconductors [11–13] and the study of metal–insulator transitions and strongly correlated quantum materials [14–17]. In s-SNOM, the metal-coated tip of an atomic force microscope (AFM) is illuminated with monochromatic IR light and produces a tightly confined and strongly enhanced IR hot spot at the tip apex. When brought in proximity to a sample surface, the tip interacts with the sample via near-field coupling. Recording the tip-scattered light while the sample is being scanned yields a map of the local dielectric properties of the sample with nanoscale spatial resolution that is independent of the IR wavelength [18,19]. Interferometric recording of tip scattered light yields amplitude and phase images, revealing local refractive and absorptive properties [20]. Because the tip-scattered near field is a relatively weak signal, s-SNOM relies on high brilliance IR sources to efficiently focus IR light on the AFM tip and to obtain sufficient signal-to-noise ratio (SNR). Development and refinement of IR light sources expanded the s-SNOM spectral range to cover molecular vibrations in the fingerprint region and many mid-IR polaritonic excitations and has thus provided the main driver in expanding the application potential of s-SNOM in the last decade. Current quantum cascade laser (QCL) technology [5,21] offers tunable, monochromatic IR emission between 1700 and 800 cm$^{-1}$ at high output power and allows for routine IR nanoimaging to map heterogeneous structure, sample composition or polaritonic excitation with high SNR.

The implementation of infrared broadband sources in s-SNOM enabled Fourier transform infrared nanospectroscopy (nano-FTIR), a derivate of s-SNOM that overcomes the diffraction-limited spatial resolution of conventional FTIR spectroscopy. In nano-FTIR, the AFM tip is illuminated with IR broadband radiation from thermal sources [22–26], difference frequency generation (DFG) [4,27–30] or synchrotrons covering the entire mid-infrared range [31–34]. The tip-scattered light is analyzed with an asymmetric Fourier transform spectrometer, yielding IR amplitude and phase spectra with the spatial resolution of s-SNOM. Particularly for soft matter such as polymers and proteins, nano-FTIR phase spectra match well with standard infrared absorption spectra [4,35]. Nano-FTIR can thus provide unambiguous material identification based on the position and ratio of molecular and phonon resonances [4–6,24,27–40], or verification of the nature of optical excitations by resolving the dispersion relation and propagation losses of polaritons [9,41–43].

The combination of s-SNOM (mapping sample heterogeneity) with nano-FTIR (material identification) offers the most comprehensive sample characterization, which is particularly beneficial in case of unknown samples. However, this poses a practical problem as both monochromatic and broadband infrared light sources need to be owned, which are expensive. Further, the spectral coverage of current QCL technology is limited in comparison to broadband infrared sources, and thus QCL-based nano-imaging becomes impractical for frequencies below 800 cm$^{-1}$ where many interesting material excitations abound [44,45]. It is thus desirable to develop a modality for spectroscopic infrared nanoimaging based on nano-FTIR. However, even with infrared supercontinuum lasers [4], which offer the highest spectral irradiance among IR broadband sources, simple sequential acquisition of spectra is still a slow imaging method, typically requiring second-long acquisition times per spectrum and multiple hours per hyperspectral data set [6]. As the spectral irradiance of IR broadband laser technology continues to improve, it is the scanning mechanism of nano-FTIR that is increasingly becoming a bottleneck for imaging speed, i.e. the piezo stage of the Michelson interferometer that needs to be scanned repeatedly for each spatial location to acquire an interferogram.

Recently, methods for nano-FTIR imaging at significantly reduced time scales were presented. Johnson et al. demonstrated spectroscopic nanoimaging of organic mineral interfaces, where the approach of a rotating frame from nuclear magnetic resonance (NMR) was adopted to sample vibration resonances at a reduced number of interferogram points [46]. Alternatively, compressive sensing has been implemented for spectroscopic nanoimaging of phonon resonances with synchrotron radiation, where the assumption of sparsity in the spectral domain was utilized to recover the nano-FTIR signal from a subsampled data set [47]. These first studies are encouraging and point to the potential of advanced spectral encoding algorithms for speeding up spectroscopic nanoimaging. Yet, reported acquisition times (few hours) have not yet reached those of monochromatic s-SNOM imaging with QCLs (few minutes). The practicality of spectroscopic nanoimaging will critically depend on performance metrics such as acquisition times, SNR and the degree of sample and interferometer drift, the limits of which remain to be explored for nano-FTIR.

Here, we have developed broadband synthetic optical holography (broadband SOH) for rapid spectroscopic nanoimaging with nano-FTIR. To this end, we combine (i) interferometric bandpass sampling to significantly reduce the number of measurements and (ii) interferogram recording in form of a single synthetic hologram, from which the full spectroscopic data set can be reconstructed. We demonstrate broadband SOH by spectroscopic nanoimaging of molecular resonances in a polymer test sample, yielding maps of 3,800 spatial locations at 8 frequency points within an effective imaging time of 20 minutes. We evaluate SNR and obtain a precision in phase measurement in the range of 20 to 200 mrad (~1°-10°), which is sufficient to resolve typical phonon resonances and moderately strong molecular resonances.

**Broadband SOH**

We illustrate and demonstrate broadband SOH by implementing it in a commercial near-field microscope (NeaSNOM, Neaspec GmbH, Munich) using the nano-FTIR module (Fig. 1). We first show that the medium spectral bandwidth of OPO/DFG sources (a few 100s of cm$^{-1}$), as used in nano-FTIR, provides opportunity for bandpass sampling [48] of the interferogram, i.e. sampling the interferogram at a rate below Nyquist. Such bandpass sampling can be applied to significantly reduce the number of interferogram points and thus speed up data acquisition by more than a magnitude compared to conventional nano-FTIR. To this end, the DFG source of the nano-FTIR module was set to emit in the spectral range from 1000 to 1400 cm$^{-1}$ (range 'B') at 350 µW spectrally integrated power. As near-field probes, we employed metalized atomic force microscope (AFM) cantilevers (PPP-FMAu, NanoWorld Ag, Switzerland) that were vibrated vertically at 40 nm amplitude and 56 kHz frequency. The infrared detector signal was demodulated at the second harmonic of the tip vibration frequency to suppress background contributions.

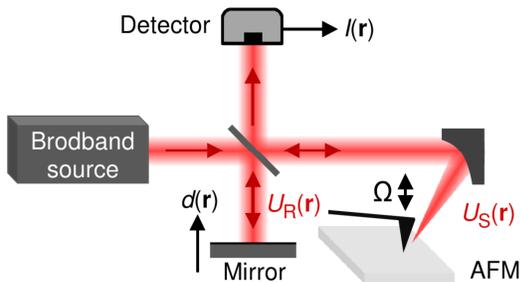

**Figure 1: Spectroscopic nano-imaging with broadband SOH implemented in nano-FTIR.**

Figure 2(a) shows a typical nano-FTIR interferogram recorded on silicon, a spectrally flat test sample. The Fourier transform (FT) of the interferogram revealed that signal content was limited to a narrow frequency band close to zero (DC) frequency (Fig. 2(b), plot (i)). This is because the interferogram was sampled at high rate as typical in nano-FTIR, that is at intervals of $\Delta d = 0.4$ μm or at 4.5 times the Nyquist rate with respect to the highest frequency (1400 cm$^{-1}$). This form of sampling is referred to as oversampling. We simulated interferogram sampling at larger intervals $\Delta d$ by decimation of the original interferogram and calculated the FT (Fig. 2(b)). We empirically found that interferogram sampling at $\Delta d = 5.4$ μm made optimal use of the spectral bandwidth, while also providing separation between the near-field spectrum and its complex conjugate (Fig. 2(b), plot (vi)). Thus, aliasing was avoided and correct reconstruction of the near-field spectrum was possible (red curve in Fig. 2(c)), as it was verified by comparison with the spectrum of the original interferogram (black curve in Fig. 2(c)). This modality of sampling is referred to as bandpass sampling. Importantly, bandpass sampling resolves the near-field spectrum with the same number of frequency points (red circles) as oversampling (black circles) within the spectral range of the source (white area in Fig. 2(c)). Thus, the reduction in interferogram points from 230 to 17 points promises a factor 13 faster imaging at no loss in spectral resolution. This reduction is valid under the assumption that the IR source is previously known to be of limited spectral bandwidth. We note that our source exhibited low and high frequency tails outside of the assumption of a bandlimited source (indicated by the grey areas in Fig. 2(c)), that folded back on to the center of the spectrum and yielded slight distortion, which could explain the discrepancy at lower wavenumbers (Fig. 2(c), near 1000 cm$^{-1}$). These spectral tails of the source could be removed by optical filtering of the laser beam with infrared bandpass filters.

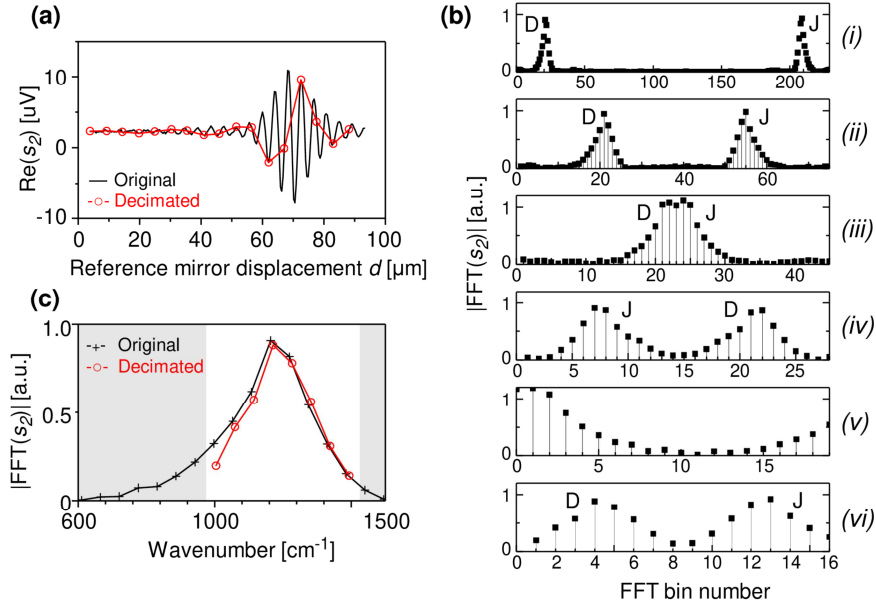

**Figure 2: Bandpass sampling as a strategy to significantly reduce the number of interferogram points at same spectral resolution.** (a) nano-FTIR interferogram on Si. (b) Fourier transform of the interferogram in (a) when resampled at different rates: (i) original sampling interval $\Delta d = 0.4\,\mu m$, (ii) $\Delta d = 1.2\,\mu m$, (iii) $\Delta d = 2.0\,\mu m$, (iv) $\Delta d = 3.2\,\mu m$, (v) $\Delta d = 4.6\,\mu m$ and (vi) $\Delta d = 5.4\,\mu m$. Labels 'D' and 'J' mark direct and conjugate spectra, respectively. (c) FT of the original (black, $\Delta d = 0.4\,\mu m$) and decimated (red, $\Delta d = 5.4\,\mu m$.) interferograms in (a), validating spectral fidelity for interferogram bandpass sampling. Acquisition parameters: 90 μm total interferogam length, 230 pixel, 10 ms integration time, 32-fold spectral averaging.

Direct implementation of bandpass sampling in conventional nano-FTIR spectroscopy is impractical because of the slow settling time of long travel piezo stages that move the reference mirror. To efficiently leverage the dramatic speed improvement afforded by interferogram bandpass sampling, we apply interferogram recording in form of synthetic holograms. To this end, we have adapted synthetic optical holography (SOH) [49], which was originally introduced for rapid nano-imaging in s-SNOM using monochromatic lasers, and later for quantitative phase imaging in other scanning microscopy methods such as confocal microscopy [50,51] and scanning cavity microscopy [52,53]. SOH uses synthetic reference waves to encode the complex field scattered from the tip in a single hologram image, from which amplitude and phase images can be reconstructed with Fourier transform (FT) filtering. SOH allows for the generation of synthetic reference waves with complex and arbitrary patterns that have no direct analog in classical holography, which we exploit here to encode nano-FTIR interferograms across the image. Specifically, in broadband SOH, at each position $\mathbf{r} = (x, y)$, the scattered field from the near-field probe, $U_S(\mathbf{r}, \omega)$, is analyzed with an asymmetric Michelson interferometer where the tip is located in one arm of the interferometer (Fig. 1). Translation of the reference mirror, located in the other arm of the interferometer, samples the interferogram of the tip-scattered light. In contrast to conventional nano-FTIR, the position $d(\mathbf{r})$ of the reference mirror is held constant during the acquisition of a line scan (here assumed in $x$-direction) and is moved to the next position after the completion of a line scan. Thus, while the AFM tip is rapidly scanned over the sample surface, the reference mirror is stepped through the positions specified in the decimated interferogram in Fig. 2(a) (red points) for every line $y$, which effectively encodes nano-FTIR interferograms across the image. The detector signal, $I(\mathbf{r})$, is demodulated at the $n$-th order of the tip vibration frequency, $\Omega$, to suppress background contributions and thus extract the local near field scattered by the tip. By recording the demodulated signal, $I_n(\mathbf{r})$, pixel-by-pixel, we obtain a synthetic near-field hologram with nano-FTIR. Such holograms exhibit a regular, horizontal fringe pattern that is periodic in the $y$ direction (Fig. 3(c), obtained with a polymer test sample, see below). The curly brackets in the zoom of Fig. 3(c) indicate the period length, i.e. the length of one interferogram. Importantly, the FT of the hologram shows clear term separation into numerous direct terms (labeled 'D'), conjugate terms (labeled 'J') and one autocorrelation term (labeled 'C' in Fig. 3(d)). The $m$-th direct term corresponds to the $m$-th frequency bin of the direct spectrum, as illustrated in Fig. 3(e). Thus, spectrally-resolved near-field amplitude and phase images, $A_S^m(\mathbf{r})$ and $\varphi_S^m(\mathbf{r})$, can be straight-forwardly reconstructed by (i) spatial filtering in the FT of the hologram, as exemplarily indicated by the dashed box in Fig. 3(d), (ii) recentering and (iii) performing an inverse FT for each direct term. The resulting data set spectrally resolves the tip-scattered field, $U_S(\mathbf{r}, \omega)$, at the same frequency points and with the same spectral resolution as with the bandpass-sampled interferogram (Fig. 2(c)). We note that the original implementation of SOH, that is linear-in-time movement of the reference mirror, will not work in nano-FTIR because (i) the limited coherence length of broadband sources introduces diminishing fringe contrast toward the hologram periphery [54] and (ii) continuous source spectrum would yield overlap of terms in the FT of the hologram, preventing spectroscopic resolution of near-field amplitude and phase images.

**Experimental Results**

We tested our method by imaging a polyvinyledene fluoride and hexafluoropropylene copolymer (FP) film of 100 nm thickness on a silicon substrate that contained FP particle aggregates of $\sim 100$ to 400 nm size with a slightly shifted spectral response (film absorption peak near 1,120 cm$^{-1}$ vs. aggregate absorption peak near 1,160 cm$^{-1}$, see Methods for details). Sample topography is shown in (Fig. 3(b)). Hologram acquisition was performed by stepping the reference mirror in intervals of $\Delta d = 5.4$ µm for every completed line scan, thus bandpass sampling the interferogram with the parameters previously established in Fig. (2), (vi) and with the same instrument settings. The resulting synthetic hologram exhibited a horizontal fringe pattern that repeated every 17 lines (Fig. 3(c)). FT of the hologram reveals eight discrete direct terms in the lower half space (Fig. 3(d), labeled 1 – 8), which correspond to center frequencies between 1000 and 1400 cm$^{-1}$ as indicated in Fig. 3(a). Near-field amplitude and phase images (Fig. 4(a)) were reconstructed from the FT of the hologram (Fig. 3(d)), as described above, spatially and spectrally resolving the different sample constituents. We note that we applied 6 times *spatial* oversampling in hologram recording in the slow scanning direction (y-axis) to partially compensate the reduced spatial resolution in the reconstructed images (see supplementary note 1).

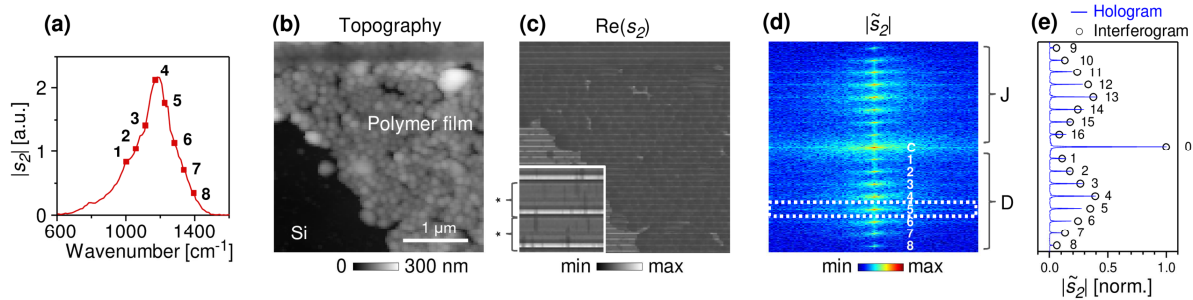

**Figure 3: Broadband SOH enables spectral imaging with nano-FTIR.** (a) Bandwidth of the DFG source of the nano-FTIR module. (b) Topography of a polymer test film on silicon (Si). (c) Synthetic hologram acquired with the acquisition parameters as in Fig. 2(a), red points. Inset shows a digital zoom, revealing the hologram fringe pattern. The curly brackets (asterisk) next to the zoom mark the length of one bandpass sampled interferogram that is repeated in y-direction. (d) FT of the hologram, revealing term discretization and separation (plotted in logarithmic color scale). Labels 'D' and 'J' mark direct and conjugate terms, respectively. Direct terms are further labeled 1 – 8 in correspondence to the frequencies marked in (a). (e) Correspondence of the frequency bins of the bandpass-sampled interferogram in Fig. 2(b), (vi) (circles) with the terms in the FT of hologram (vertical line out across center of (d)). Note that hologram (c) and FT (d) (100 x 612 pixel) are displayed with square aspect ratio to reflect the real scan dimensions 3 µm x 3 µm.

We next demonstrate the analytical capabilities of spectroscopic nanoimaging with broadband SOH (Fig. 4(a)). Spatial features on the sample were recorded with 30 nm pixel size across an area of 3 μm x 3 μm and spectral features with 55 cm$^{-1}$ resolution in the range of 1000 to 1400 cm$^{-1}$. This capability allowed for clear identification of the FP polymer film based on molecular absorption observed as a peak in the near-field phase of ~0.6 rad at 1226 cm$^{-1}$. The FP particle aggregates exhibited stronger response in phase (~2 rad) and amplitude (~1.2) peaking at 1170 cm$^{-1}$ and 1114 cm$^{-1}$, respectively. The measured absorption bands were reproduced by nano-FTIR spectroscopy in peak position, peak height and peak shape, as revealed by direct comparison at selected locations on the sample (Fig. 4(b)). This confirmed that broadband SOH provided accurate near-field spectral data. Fig. 4(c) quantifies noise levels in the near-field phase at regions of the FP film and the Si substrate, which are representative for typical low and high refractive index materials, respectively. We observed that the standard variation of the near-field phase generally scaled with the input power of the broadband source and the near-field amplitude response of the material. Thus, nano-imaging with phase sensitivity of ~20 mrad (~1°, on Si) and ~40 mrad (~2°, on FP polymer film) was possible near the center of the source emission. Lowest (1) and highest (8) frequency bins yielded higher phase noise and are thus not shown here. We provide further validation of broadband SOH in a simulation in supplementary note 2.

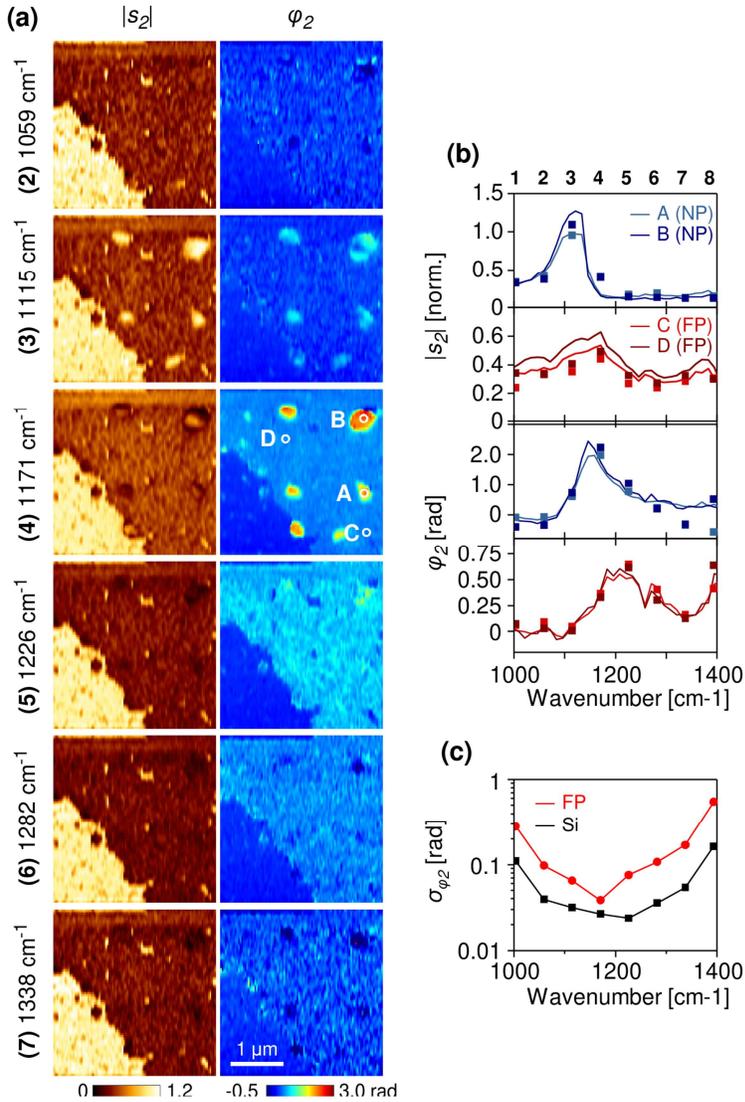

**Figure 4: Demonstration of rapid spectroscopic imaging with a polymer test sample.** (a) Spectral near-field amplitude and phase images reconstructed from the hologram in Fig. 3(c) and normalized to silicon. (b) Comparison of near-field phase as obtained from the spectral data set in (a) with nano-FTIR spectra, both taken at locations A-D. (c) Standard deviation of the near-field phase as obtained from regions of FP (fluorine copolymer) and Si (silicon).

**Discussion**

Broadband SOH is based on interferogram bandpass sampling and thus offers opportunity to trade signal-to-noise ratio for high imaging speed, which cannot be easily done in conventional nano-FTIR imaging. In the presented experiment, we assumed a typical imaging scenario in s-SNOM where field of view (9 $\mu m^2$), spatial (30 nm) and spectral resolution (55 $cm^{-1}$) are sufficient to locate and identify many molecular and phonon resonances. While spectroscopic nanoimaging with conventional nano-FTIR can be expected to take hours, our experiments showed that such spectroscopic imaging can now be done on the time scale of 20 minutes. To our knowledge, this is the fastest demonstration of spectroscopic nano-imaging with a broadband source to date and it approaches the performance of near-field imaging with QCL.

Further, broadband SOH provides two practical advantages over previous methods for spectral nano-imaging. First, spectral imaging with broadband SOH only requires the acquisition of a single hologram, which is technically simple and fast. Particularly, the overhead of conventional nano-FTIR imaging is avoided, i.e. the repeated re-centering of the sample (to correct for sample drift) and re-measuring of reference spectra (to compensate for interferometer drift) during data acquisition [6]. Secondly, sample drift does not affect spectral fidelity as long as this drift is small across the length of one interferogram (here, 17 lines of the hologram), which is easy to achieve. In comparison, previous methods for rapid spectroscopic imaging (e.g. ref. [46]) sampled the interferogram on an image-by-image basis, where each near-field image was acquired for a fixed position of the reference mirror, which requires correction of sample drift.

Finally, broadband SOH is well suited for rapid imaging of a single narrowband spectral feature because, in this case, high subsampling ratios can be achieved and hence bandpass sampling becomes particularly efficient. The needed narrowband sample illumination can be provided either directly using high-power IR sources with medium-sized spectral bandwidth [55] or by employing infrared bandpass filters – commercially available from several vendors – to reduce the source bandwidth of IR supercontinuum or synchrotron sources. Importantly, the spectral resolution of broadband SOH is defined by the length of the interferogram as in nano-FTIR, and hence we expect that a spectral resolution of ~6 $cm^{-1}$ can be reached [24]. In principle, our method can be extended to cover several disjoint regions in the IR spectral range, which would be useful for probing multiple absorption bands at very different frequencies simultaneously. To this end, the source spectrum could be tailored to cover just a narrow region around each absorption band (e.g. with optical bandpass filters) and bandpass sampling would need to be adjusted such that all individual regions spectrally separate in the FT of the subsampled interferogram, which can be tested with the procedure illustrated in Fig. 2.

## Conclusions

We have presented broadband SOH for rapid spectroscopic nanoimaging with nano-FTIR that is based on interferogram bandpass sampling combined with synthetic optical holography. In particular, we demonstrated spectroscopic nanoimaging of vibrational resonances on the order of tens of minutes rather than hours, which could find application in routine imaging of phonon resonances and molecular resonances that yield a moderately large near-field phase response. Broadband SOH provides a practical alternative for QCL-based nanoimaging in cases where (i) only the nano-FTIR module is available for use or (ii) nanoimaging outside the spectral range of QCLs is desired. This capability may enable the direct observation of polaritons in atomically thin materials and semiconductors, which abound the long wave IR spectral region below 800 cm$^{-1}$ [44], and other low-energy excitations in solids [45], where coverage by current QCL technology is sparse. Our method makes use of existing hardware (DFG laser, nano-FTIR module) and can be implemented in software. Beyond nano-FTIR, our method could also be implemented in other scanning methods including FTIR microspectrometry and spectroscopic LIDAR where the combination of an interferometer and a one-pixel detector could provide the function of a classical spectrometer and an array detector.

**Competing interests:** RH is a co-founder of Neaspec GmbH, a company producing scattering-type scanning near-field optical microscope systems such as the one used in this study. MS, RH and PSC are authors of US patent 9,213,313. The other authors declare no competing interests.

**Acknowledgements:** M. S. acknowledges support by the European Union's Horizon 2020 research and innovation programme under the Marie Sklodowska-Curie grant agreement No 655888 (SYNTOH). The authors further acknowledge financial support from the Spanish Ministry of Science, Innovation and Universities (national project RTI2018-094830-B-100 and the project

MDM-2016-0618 of the Marie de Maeztu Units of Excellence Program) and the Basque Government (grant No. IT1164-19).

**Methods**

<u>Sample preparation</u>

Test sample was prepared using a water-based polymer dispersion (latex) of polyvinyledene fluoride (PVDF) and hexafluoropropylene (HFP) copolymer (95/5 % in weight) made by emulsion polymerization, kindly supplied by The Chemical Engineering group of the University of Basque Country-POLYMAT. The final particle size of the latex was around 129 nm measured by Dinamic Light Scattering (DLS, Malvern Nanosizer). The latex was diluted at 0.2% solids content and mixed under stirring. Then, the sample was cast onto silicon wafer through spin coating and dried overnight at room temperature. The film yielded two different conformations consisting of a polymer film formed by particles that have coalesced and individual aggregates formed by particles that have not coalesced, as a result of the specific drying conditions. Nano-FTIR measurements confirmed that both conformations yielded slightly different spectral signatures, which are detected and imaged in the experiments in the main text: molecular absorption near 1,220 $cm^{-1}$ for the film and near 1,170 $cm^{-1}$ for the aggregates.

# Rapid infrared spectroscopic nano-imaging with nano-FTIR holography

**Supplementary Material**

Martin Schnell[1]*, Monika Goikoetxea[1], Iban Amenabar[1], P. Scott Carney[2], and Rainer Hillenbrand[3,4]

1 CIC nanoGUNE BRTA, 20018 Donostia-San Sebastián, Spain

2 The Institute of Optics, University of Rochester, 480 Intercampus Drive, Rochester, NY 14620

3 IKERBASQUE, Basque Foundation for Science, 48013 Bilbao, Spain

4 CIC nanoGUNE BRTA and UPV/EHU, 20018 Donostia-San Sebastián, Spain

* schnelloptics@gmail.com

**Supplementary Note 1 – Acquisition time and oversampling considerations for broadband SOH**

Ideally, broadband SOH would be implemented in hardware where scanning of the sample and stepping the reference mirror are performed simultaneously, allowing for rapid hologram acquisition. However, the current s-SNOM system did not support such a modality and hence the hologram was acquired as individual line scans using the microscope's script language. This special implementation yielded unnecessary overhead that could be avoided in the future by a hardware or more low-level software implementation. In the presented experiment, total imaging time was 37 minutes for a 100 x 612 pixel hologram at 10 ms integration time, of which 20 minutes were actual recording time – including forward and backward trace – and 17 minutes of overhead related to implementing hologram recording using script language. The bulk of this overhead time was spent on repeated but unnecessary tip movement before each line scan.

The reconstructed near-field images contained a nominal number of spatial locations of 100 x 612 and are displayed with a square aspect ratio in Figs. 3,4 of the main text to reflect the real scan dimensions. Note that the number of lines (612) were chosen to be an integer multiple of the number of interferogram points to avoid artifacts in the hologram processing. Taking into account the six-times spatial oversampling in $y$ in hologram recording and the filter width (1/16 in $y$) in the reconstruction process, the effective number of spatial locations is estimated as 100 x 38 (=3,800). Nevertheless, the image quality appeared better than expected considering this conservative value. We attribute this observation to the fact that spatial oversampling in SOH provides better localization of morphology features (e.g. polymer film edge). A possible application of such high spatial oversampling could be the detection of small, sparsely distributed objects in large-area scans. This is because in broadband SOH, the tip raster-scans the sample in very fine steps in $y$ (in our experiments, the line step is 5 nm). In comparison, the tip raster-scans the sample only coarsely in $y$ direction, when conventional nano-FTIR imaging is applied that determines spectral information on a pixel by pixel basis.

Generally, in broadband SOH, the degree of spatial oversampling is determined either by the inherent bandwidth of the sample or the bandwidth of the image passed by the optical system (given by the tip radius in s-SNOM), whichever is smaller. A practical implication is that the oversampling requirement can be relaxed in many cases because some degree of spatial oversampling is almost always applied in s-SNOM, i.e. it is typically intended to resolve morphological features across two or more pixels. In the main text, we reduced this requirement by a factor of 3 by applying 6-times oversampling (rather than 16-times = inverse filter width), still resolving the relevant details of the polymer sample (Fig. 4). Larger objects with only low spatial frequency content may allow for even higher reduction factors. This reduction promises an additional speed advantage over non-holographic methods such as conventional nano-FTIR spectral imaging which determine spectra on a pixel-by-pixel basis.

## Supplementary Note 2 – Simulation of Broadband SOH

To further validate the precision of our method, we performed a simulation of broadband SOH using a complete 3D interferogram data set of a polymer mix sample [1]. Particularly, we show that holographic encoding of interferograms allows for accurate retrieval of near-field spectra. This approach allowed for a direct signal comparison between broadband SOH and nano-FTIR because experimental variability between separate acquisitions of broadband SOH and nano-FTIR data sets could be avoided (e.g. detector noise and interferometer drift). To this end, we first trimmed interferogram length of the original 3D interferogram set to about 90 μm (170 interferogram points) to reduce spatial resolution to 54 cm$^{-1}$ and thus match the experimental conditions to those in the main text. We then estimated the subsampling parameter and found that interferogram sampling at only 17 points made optimum use of the Fourier space (Figs. S1(a,b)). The original data set contained 62 x 90 spatial locations. We generated a synthetic hologram by concatenating interferograms in the $y$ direction to produce a single image of 62 x 1,530 pixels (Fig. S1(d)). More precisely, the subsampled 3D data set $(x, y, d)$ was re-mapped to hologram coordinates $(x, (y-1) * 17 + d)$, where $x$, $y$ are indices of spatial position and $d$ is the index of reference mirror position. Filtering of the term in Fourier space (Fig. S1(e)) yielded a set of near-field phase images (Fig. S1(g)). Comparison with conventional nano-FTIR spectroscopic imaging showed identical image contrast, as further quantified by comparing spectra (Fig. S1(f)). Slightly larger noise is apparent with the data reconstructed from the hologram as a result of the reduced number of interferogram points and thus shorter effective recording time of the entire interferogram.

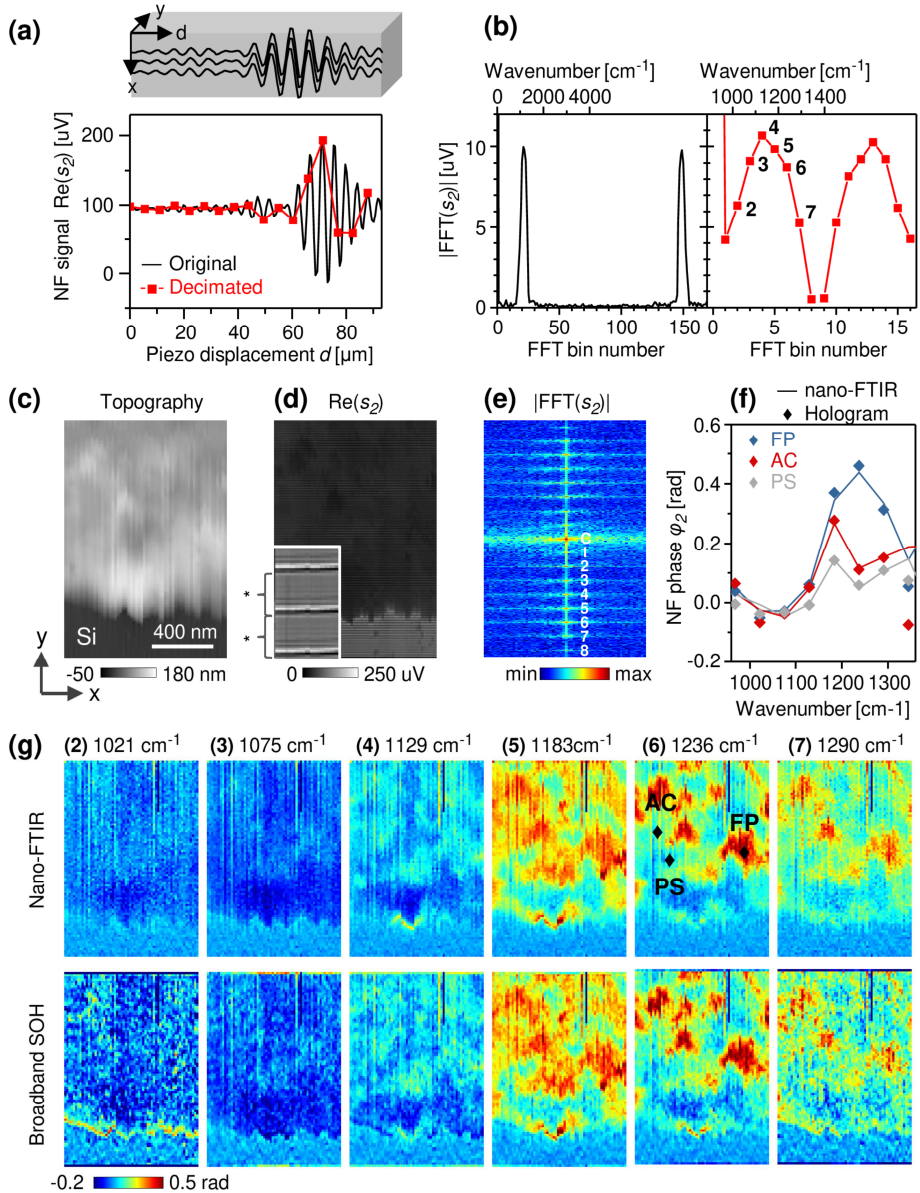

**Figure S1: Validation of broadband SOH by simulated spectral nanoimaging**. (a) Example interferogram of the original 3D nano-FTIR data set. (b) FT of the interferogram of the original (black) and decimated interferogram (red). (c) Topography of a polymer blend consisting of fluorine copolymer (FP), acrylic copolymer (AC) and polystyrene latex (PS) on silicon. (d) Simulated synthetic hologram and (e) FT. The curly brackets (asterisk) in (d) marks the length of one subsampled interferogram mapped onto the y-axis. (f) Quantitative comparison between original spectra (line) and spectra extracted from the holographic data set (symbols). (g) Spectral near-field phase images as obtained by nano-FTIR (original data) and by broadband SOH, as reconstructed by filtering in the FT of the hologram in (e).